# A method for diffraction-based identification of annealing-produced restructuring of amorphous fullerene

V.S. Neverov[1], P.A. Borisova[1], A.B. Kukushkin[1,2], V.V. Voloshinov[3]

[1]National Research Center "Kurchatov Institute", Moscow, Russia
[2]National Research Nuclear University "MEPhI", Moscow, Russia
[3]Institute for Information Transmission Problems (Kharkevich Institute) of Russian Academy of Science, Moscow, Russia

A method is suggested for estimation of structural properties of amorphous fullerene and its derivatives produced by vacuum annealing. The method is based on the fitting of the neutron or x-ray powder diffraction patterns for scattering wave vector's modulus in the range from few units to several tens of inverse nanometers. The respective inverse problem assumes that the structured component of a sample can be described with a limited number, $N_{str}$, of candidate $sp^2$ carbon structures (fullerenes, flat and curved flakes with graphene-like atom arrangement) of a limited number of atoms, $N_{atom}$. These structures are packed heterogeneously, in the domains with various average density of atoms and various degree of ordering of structures, using the Rigid Body Molecular Dynamics with variable parameter of pair interaction of atoms in the neighboring rigid-body nanostructures. The method is applied to interpreting the data of neutron diffraction by an amorphous fullerene annealed at 600, 800, 850, 900 and 1000 $^{o}$C. The results for $N_{str} = 36$ and $N_{atom} = 14 \div 285$ enabled us to quantify structural properties of the samples in terms of the average size and curvature of the $sp^2$ carbon structures, and analyze sensitivity of results to the layout of these structures in the domains (mixture of various structures in each domain vs. mixture of domains of identical structures).

## 1. Introduction

Identification of structural contents of amorphous materials in the nanometer length scales from an analysis of powder diffraction patterns (x-ray, neutron, etc.) is known to be a very complicated task: the lack of spatial ordering typical to crystalline media leaves much freedom in the reconstruction of structural content [1]. In the case of a single or few structural blocks and spatial homogeneity of the blocks in the sample one can use the method of the Reverse Monte Carlo (RMC) modeling (cf., e.g., [2]). The RMC reconstruction is based on the fitting of the experimental pair distribution function of relative atomic positions with that calculated for variable (unknown) layout of atoms in the structural block(s) and known chemical composition of the sample. In this approach the variation procedure works with the positions of separate atoms, and if the initial atomic ensemble is random (ab-initio approach) the presence of chemical bonds between them may be recognized only on the final step of the algorithm (cf., e.g., identification of fullerenes in [3]).

There is an alternative to the RMC approach which is based on the variation within a set of many predefined nanoscale structural blocks (of course, of limited total number). Here the variety of possible chemical bonds is introduced on the first step of algorithm. The remaining freedom of packing these structural blocks in the sample requires additional numerical modeling which may generate various samples as various snapshots of an ensemble of blocks during the rigid-body dynamics of blocks subject to inter-atomic forces between them. In this approach, both the chemical bonds and the atomic positions within certain block are fully conserved.

A part of the outlined method (namely, without full variation of packing) has been used in the interpretation [4] of measurements of the synchrotron XRD by hydrocarbon films deposited in the vacuum vessel of tokamak T-10 in Kurchatov Institute. This enabled us to recover the structural content of carbon nanostructures of various topology and size in these films. To solve the problem [4] we developed the following software and methods: (i) software package

XaNSoNS (X-ray and Neutron Scattering on Nanoscale Structures) for calculating the diffraction patterns of the ensembles of nanostructures, (ii) method [5] of an approximate description of the position of carbon atoms in a curved graphene sheet, (iii) optimization procedure and the respective RESTful web-services [6] created in the Mathcloud distributed environment [7] for fitting the experimental x-ray and/or neutron diffraction patterns with a set of numerically simulated ones. In this work we use the tools of items (i) and (iii).

Here we develop and apply a method of diffraction-based identification of structural content for particular case of amorphous fullerene and its derivatives produced by vacuum annealing, which is of interest for various applications. The respective inverse problem assumes that the structured component of the samples can be described with a limited number, $N_{str}$, of candidate $sp^2$ carbon structures (fullerenes, flat and curved flakes with graphene-like atom arrangement) of a limited number of atoms, $N_{atom}$. These structures are packed heterogeneously, in the domains with various average density and degree of ordering. We use the Rigid Body Molecular Dynamics (RBMD) with variable parameter of pair interaction of atoms in the neighboring rigid-body nanostructures. The RBDM technique was applied to packing of various $sp^2$ and $sp^3$ carbon structures in [8] to analyze the impact of packing on the x-ray diffraction data analysis. The model developed here is applied to interpreting the experimental data (Sec. 3) of neutron diffraction by an amorphous fullerene annealed in a vacuum at 600, 800, 850, 900 and 1000 $^o$C [9]. To interpret experimental results (Sec. 4) we apply the model for $N_{str} = 36$ and $N_{atom}$ in the range from 14 to 285. This enabled us to quantify the structural properties of the sample in terms of the average size and curvature of the $sp^2$ carbon structures, and analyze sensitivity of results to the layout of these structures in the domains (mixture of various structures in each domain vs. mixture of domains of identical structures).

## 2. Method of diffraction-based identification

The method is aimed at estimation of structural properties of amorphous fullerene and its derivatives produced by vacuum annealing. It may be represented as the following sequence of actions.

**(a)** Definition of the structural blocks which we choose for quantitative description of a structured carbon component in the sample. Particular conditions of the production of the sample and the data from other diagnostics (chemical composition of the sample is supposed to be known) may help to limit the set of possible structural blocks, e.g., to a set of a limited number, $N_{str}$, of the $sp^2$ carbon structures in particular case of this work.

**(b)** Numerical modeling of the possible layout of structural blocks within a sample. The presence of a non-structured component in amorphous materials provides some freedom in packing the structural blocks within the sample. Under these conditions numerical modeling should generate various samples as various snapshots of an ensemble of blocks during the rigid-body dynamics of blocks subject to inter-atomic forces between the blocks (the principles of numerical modeling are described below). The variation of these samples is described by continuous parameters like average density of atoms in the structured component of the sample, and degree of ordering of blocks mutual orientation, e.g. random or ordered. The variation of possible packing on a grid of the values of these parameters for *arbitrary* variable composition of structural blocks gives a number of variants which is too large even for supercomputer simulations. We suggest to restrict our analysis to the following two limiting cases.

(b1) First, we consider the case of a packing in the form of domains which contain only identical structural blocks and possess quasi-homogeneous density of atoms. Such a sample appears to be a heterogeneous material in which variation of structural content is limited to variation of the composition of domains with identical blocks (we call these as mono-structure domains). The linear size of the domain has a lower limit of 10 nm to avoid the impact of the finite size of the domain on the diffraction patterns in the range of scattering wave vector's modulus under consideration (from few units to several tens of inverse nanometers). For such domains the effects of interference of atoms, located closely to the boundaries of domains, are small, and the diffraction pattern is merely a sum of those for separate domains. The average density of atoms in the domains and (the defined below) degree of ordering of the structural blocks in the domains are the independent variables. The total number of the mono-structure domains is equal to the product $N_{str} \cdot N_p$, where $N_p$ is a number of variants with different density and degree of ordering of blocks mutual orientation in each mono-structure domain. The case of composing a sample from mono-structure domains enables one to start with a computationally feasible task and to find the structural content which may be used as a test distribution for other partial cases of the general inverse problem.

(b2) Second, we consider the case of the mixture of different structural blocks within domains (we call these as multi-structure domains) with the atomic fraction of blocks found in the case of mono-structure domains. The average density of atoms in the domains and degree of ordering of the blocks in the domains are varied similarly to the case of mono-structure domains. The total number of the domains in this case is equal to $N_p$. If the simulated scattering intensity in the case of multi-structure domains may fit the experimental diffraction pattern almost as well as it does in the case of mono-structure domains one can claim that the diffraction patterns is not sensitive to the type (mono- or multi-structure) of the domains and characterize the sample in terms of some averaged values of major parameters, e.g., average size and curvature of the $sp^2$ carbon structures for amorphous fullerene and its derivatives produced during the annealing.

If the combinations of multi-structure domains fail to fit the experiment, one has to try the cases which are close to the case of mono-structure domains. The choice of the test distributions may be suggested by particular features of the sample and its production. Here we restrict ourselves only to indication of the necessity to try such a way.

To complete the description of the algorithm we present a brief description of simulations of the packing and of the inverse problem formulation.

We use the Rigid Body Molecular Dynamics (RBMD) approach [10] to model the packing of the carbon structural blocks in the domains. In this approach, blocks are considered as rigid bodies and chemical bond formation and breakage, as well as bond angle change, is prohibited. We use the Lennard-Jones potential to describe the pair interaction between the atoms in the neighboring rigid blocks in the ensemble:

$$U(r_{ij}) = \varepsilon \left[ \left( \frac{R_{opt}}{r_{ij}} \right)^{12} - 2 \left( \frac{R_{opt}}{r_{ij}} \right)^6 \right], \qquad (1)$$

where $r_{ij}$ is the distance between the $i$-th and $j$-th atoms of different blocks (relative positions of atoms of the same block are strictly fixed), $R_{opt}$ is the minimum point, $\varepsilon$ is the potential well depth. The cut-off length is equal to 3 $R_{opt}$.

The pair potential is unable to describe, e.g., all types of layer stacking in crystalline graphite [11], but in our work we are studying the diffraction patterns of disordered material so that we can neglect the subtle effects. The average distance between the blocks in the ensemble depends on the $R_{opt}$ parameter which determines the minimum of total potential energy of the ensemble (this energy is defined as a double sum of Eq. (1) over $i$ and $j$).

The steady-state positions of the blocks in the ensemble are found as a stationary solution of the dynamic problem and correspond to the minimum of the total potential energy (of course, a local minimum of a function of all atomic positions). The domain with the layout of structural blocks in the steady-state of the ensemble is defined as an ordered domain whereas the initial state with random distribution is defined as a disordered domain. We also consider the case of a weakly ordered domain, which corresponds to an intermediate state of the ensemble during its time evolution, namely that with the potential energy equal to a half of the total potential energy in the disordered and ordered states.

To speed up the calculations, we introduced a sink of kinetic energy assuming that the blocks are moving in a viscous medium. Also, we use the quaternion representation to treat the rotation of a rigid body [12]. To reduce the computational time we put all the structures into a 3D uniform grid, the size of the cell is equal to the cut-off length of the Lennard-Jones potential so that only the atoms within the current and neighboring cells can interact, and there is no need to compute all the distances $r_{ij}$. The time integration procedures are performed in the following order: grid update, calculation of forces, angular and linear momentum update, quaternion update, relative atomic positions and center of the mass position update, grid update. The created numerical code is parallelized using the MPI.

We illustrate the formation of domains with all three types of ordering of blocks within the domain on the example of domains with a single structural block (mono-structure domain) taken as the flat $sp^2$ carbon structure of 142 atoms with graphene-like atom arrangement ($C_{142}$). The ensemble of the $C_{142}$ structures composed as an initial state for the RBMD modelling is shown in Fig. 1a. Originally the $C_{142}$ structures are randomly distributed inside a finite volume. It's not possible to distribute them with a high enough density. Therefore, in order to increase the density of the ensemble but keep the layout of the blocks random we start the modelling with the prohibited rotational motion of blocks. For each block we introduce the initial momentum directed to the center of mass of the entire ensemble with the magnitude proportional to the distance from the position of the block to the center of mass of the entire ensemble. The latter allows the blocks to form a single domain - rather than a few separate ones - of a pretty homogeneous atomic density on the space scale of few nanometers. Such a domain of the $C_{142}$ structures is shown in Fig. 1b. The density of the ensemble is almost four times higher than that in the initial state "a". The state "b" is an example of the disordered domain of $C_{142}$. At the next stage of modelling we start from the state "b" with an allowed rotation of blocks. Fig. 1d. shows the steady-state of the ensemble of $C_{142}$ structures. For this particular case the potential energy of the ensemble in the steady-state "d" is almost four times lower than that in the state "b". The $C_{142}$ structures form a graphitized clusters with the number of layer from 2 to 12. This illustrates why we call the layout of blocks in the steady-state of the ensemble an ordered. This state is an example of the ordered domain of $C_{142}$. Fig. 1c shows the ensemble of $C_{142}$ structures in the state in which the total potential energy of the ensemble is a half of the sum of total potential energies in the "b" and "d" states. In this state some of the $C_{142}$ structures form two- or three-layered clusters while some others didn't. This state is an example of the weakly-ordered domain of $C_{142}$. The $R_{opt}$ parameter value for each pair of structures in the ensemble in Fig. 1 is set in the range from 0.35 nm to 0.4 nm with an average value of 0.375 nm within the ensemble.

The difference of layout of blocks in the disordered, weakly ordered and ordered domains becomes smaller with an increase of the curvature of structural blocks (namely, $sp^2$ carbon structures), averaged over domain, and is negligible for the domains of the $C_{60}$ fullerene molecules.

Of course, the value of the potential energy depends not only of the ordering of the structural blocks layout within the ensemble but also of their density. The density of the ordered domain is always higher than that of the disordered one of the same structural content. To vary the atomic density in the domains independently to the degree of ordering in the domains we take the $R_{opt}$ parameter in Eq. (1) a variable. This enables us to model the domains of different density while the defined above degree of ordering of domains may be considered as essentially the same.

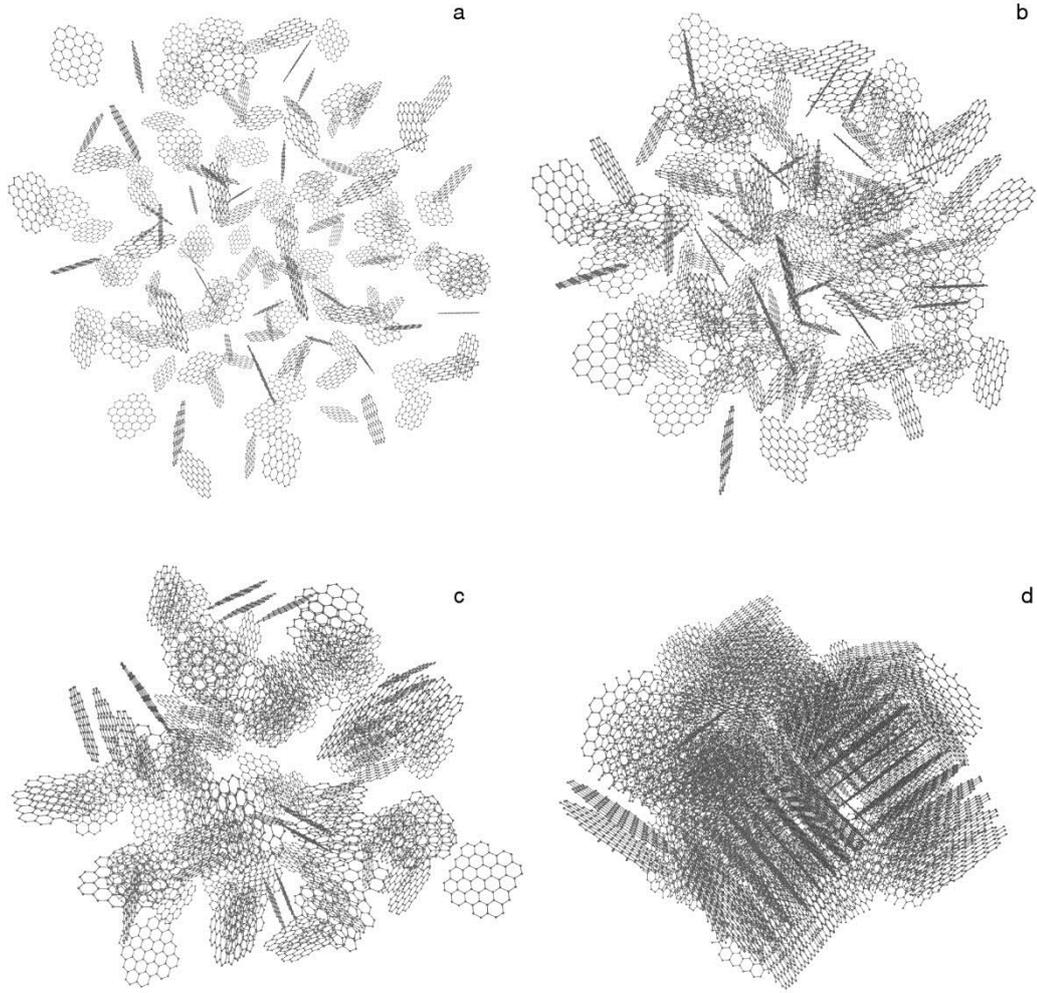

Fig. 1. The states of ensembles of the $sp^2$ carbon structures, $C_{142}$, at subsequent stages of the RBMD modelling: (a) initial random distribution in a finite volume; (b) after evolution with the prohibited rotational motion of $C_{142}$ structures (a disordered domain); (c) further, after the rotation of $C_{142}$ structures is allowed (a weakly-ordered domain); (d) steady-state (an ordered domain). The spatial scale of the picture "a" is smaller than that in "b", "c" and "d".

We work with the experimental powder diffraction pattern rather than the experimental pair distribution function (PDF) of relative atomic positions because the calculation of the PDF from a noisy experimental diffraction pattern may produce additional errors. The fitting of the experimental diffraction pattern is formulated as the minimization of the norm of the vector, each element of which represents the difference between the experimental pattern and the simulated scattering intensity at the current point of discrete finite set of the values of the scattering vector modulus $q$. The dimension of this "discrepancy vector" equals to the total number of points in the experimental data. The simulated intensity is calculated as a sum over the calculated diffraction patterns of all domains with unknown weight coefficients. The two additional unknown parameters should be included in the simulated intensity. The first one is an unknown constant background and the second one is a contribution of the non-structured amorphous component, including the carbon and impurities, to the experimental diffraction pattern. The formula for the vector element is as follows:

$$Z_j(\mathbf{x}, b) \equiv S_{exp}(q_j) - \sum_{i=1}^{N} S_i(q_j) x_i - a S_p(q_j) - b \quad (j = 1{:}m) \tag{2}$$

where $S_{\exp}(q)$ is an experimental powder diffraction pattern; $S_i(q)$ is a calculated diffraction pattern for the $i$-th domain normalized to the number of atoms in the domain; $S_p(q)$ is a normalized ($\sum_{j=1}^{m} S_p(q_j) = \sum_{j=1}^{m} S_{exp}(q_j)$) contribution of the amorphous non-structured component, including the carbon and impurities, to the experimental diffraction pattern (the problem of identifying this in a particular case is addressed in Sec. 4); $j$ is the number of the current point in the discrete space of scattering vector modulus $q$; $N$ is the total number of the domains ($N = N_{str} \cdot N_p$ in the model of mono-structured domains and $N = N_p$ in the model of multi-structured domains); $m$ is the total number of points in the experimental data. The unknown parameters are as follows: the weight coefficients $x_i$; the weight of the contribution of the non-structures amorphous component, including the carbon and impurities, to the experimental pattern, $a$; and the constant background, $b$. The atomic fraction of the $i$-th domain in the structured component of sample is equals to $x_i / \sum_{i=1}^{N} x_i$. There are three additional constraints on the unknown parameters:

$$x_i \geq 0 \ (i = 1:N), \tag{3}$$

$$B_{min} \leq b \leq B_{max}, \tag{4}$$

where $B_{min}$ and $B_{max}$ are the lower and upper bounds for the unknown constant background in the experimental data.

We consider the following norm of the vector (2) which is to be minimized:

$$\sum_{j=1}^{m} |Z_j(\mathbf{x}, b)| \xrightarrow[\mathbf{x}, a, b]{} \min \ (L_1 \text{ norm}). \tag{5}$$

The form of the function $S_p(q)$ is not a universal one, and its evaluation depends on the specific experimental data. The procedure of the evaluation of $S_p(q)$ for the experimental data of Sec. 3 is given in Sec. 4.

The knowledge of the weight coefficient $x_i$ enables one to estimate the structural properties of the sample.

## 3. Experimental data

Initial samples of $C_{60}$ fullerenes of 99.5% purity have been produced by "NeoTechProduct" by the high-temperature treatment of graphite, followed by isolation with organic solvents and subsequent chromatographic separation. The neutron diffraction patterns of initial crystal fullerenes revealed the FCC lattice with a period a = 1.416 nm.

The amorphous fullerenes were produced from crystal fullerenes with a ball mill. The grinding bowl of 80 ml volume and five balls of 20 mm diameter were made of agate. The powder-to-ball mass ratio was 1:14. Since the mechanical milling is carried out in air, the presence of gas impurities (oxygen and nitrogen) in the unannealed sample is very likely.

The thus obtained samples were subjected to a high-temperature (up to 1000°C) step annealing in a VP1/20 vacuum electric shaft furnace (~ 5·10$^{-5}$ torr) and then cooled to room temperature, at which the diffraction experiments were performed.

The neutron diffraction analysis of the sample's structure was performed at the multidetector diffractometer DISK [13] for the 0.1668 nm wavelength of monochromatic thermal neutrons [9].

In this work we examine the samples annealed at 600, 800, 850, 900 and 1000 °C referenced as $S_{600}$, $S_{800}$, $S_{850}$, $S_{900}$ and $S_{1000}$ respectively and the unannealed sample of amorphous $C_{60}$ fullerene, referenced as $S_{25}$.

## 4. Theoretical interpretation

A comparison of the diffraction pattern of the $S_{25}$ sample with the calculated diffraction pattern for the amorphous $C_{60}$ (see Fig. 2) shows that the heights of local minima for the experimental and theoretical curves are different while their positions at x-axis are the same (see black dots in Fig. 2). This allows to assume the presence of an unknown contribution of the amorphous non-structured component, including carbon and impurities (mostly, oxygen and nitrogen) in the diffraction pattern. We assume this contribution to be a smooth function which for the experimental data from Sec. 3 may be presented in the polynomial form. In Eq. (2) the respective contribution may be taken as a cubic polynomial. Indeed, a subtraction of the smooth cubic polynomial from the experimental curve gives us a good fitting of the experimental local minima (empty dots in Fig. 2) by the theoretical curve. We assume the validity of the same polynomial representation of the contribution of a non-structured component, including the impurities, $S_p(q)$ in Eq. (2), for all samples; only the fraction $a$ can vary.

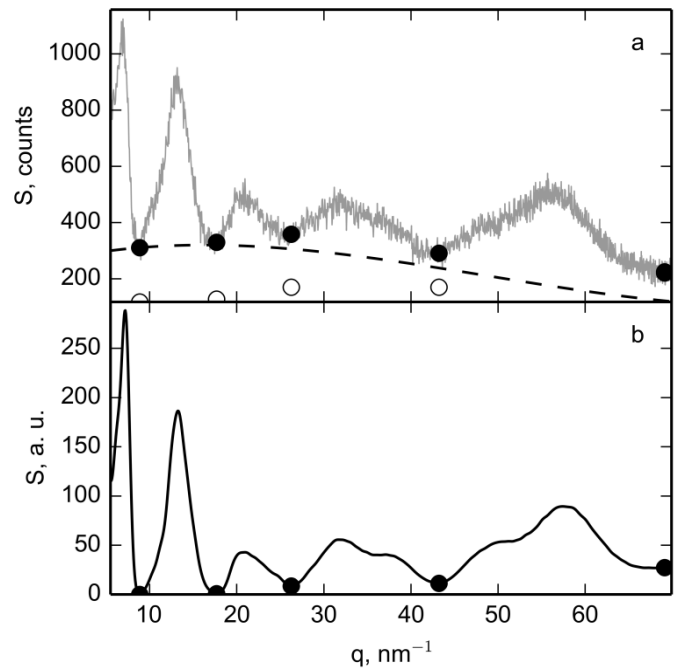

Fig. 2. Comparison of the (a) experimental (unannealed sample, grey solid) and (b) calculated (amorphous $C_{60}$ with average intermolecular distance ~ 0.33 nm) neutron diffraction patterns. Local minima are shown by the black dots. Dashed black curve is a cubic polynomial fit which, after the subtraction from experimental curve, gives the heights of local minima (empty dots) closely to those of the theoretical curve.

The set of structural blocks selected for quantitative description of a structured carbon component in the samples includes the $sp^2$ carbon structures of different number of atoms, shape and curvature. This includes the following structures of the total number $N_{str} = 36$: (i) $C_{60}$ fullerene molecules; (ii) flat graphene flakes of linear size of 1.0, 1.4, 2.2, and 2.6 nm ($C_{34}$, $C_{62}$, $C_{142}$ and $C_{200}$); (iii) fragments (the caps of different size) of large spherical fullerenes with sphere radius from 0.35 to 1.26 nm ($C_{14}$, $C_{30}$, $C_{36}$, $C_{41}$, $C_{52}$, $C_{60f}$, $C_{78}$, $C_{86}$, $C_{90}$, $C_{120}$, $C_{120f}$, $C_{125}$, $C_{129}$, $C_{143}$, $C_{160}$, $C_{185}$, $C_{208}$, $C_{208f}$, $C_{250}$, $C_{270}$, $C_{285}$); (iv) fragments (halves) of capsule-like fullerenes with capsule (tube) radius from 0.35 to 0.61 nm ($C_{45}$, $C_{58}$, $C_{63}$, $C_{73}$, $C_{80}$, $C_{95}$, $C_{119}$, $C_{147}$, $C_{153}$, $C_{195}$). The subscript indicates the number of atoms in the carbon structure. Letter "f" is added to some subscripts to denote lower curvature of the structure with the same number of atoms. Some examples of the structures of different shape are shown in Fig. 3. Representation of the curved $sp^2$ carbon structures with the fragments of large fullerenes allows to avoid the quantum chemistry calculation of the atomic layout in these structures.

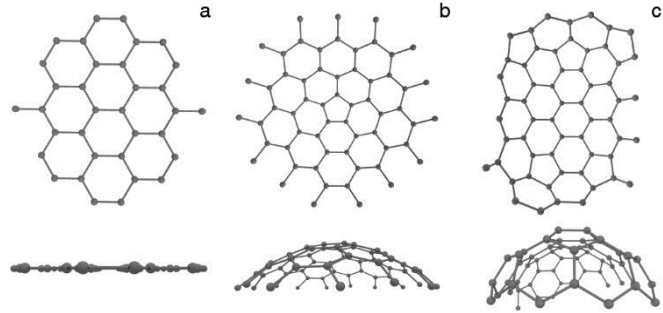

Fig. 3. The layout of atoms in the *sp²* carbon structures of different shape and curvature: (a) $C_{62}$ flat graphene flake, (b) $C_{60f}$ fragment of large spherical fullerene with sphere radius 0.82 nm, (c) $C_{58}$ fragment (half) of capsule-like fullerene with capsule (tube) radius 0.43 nm. Top row gives a top view (for b and C, from the concave side of the structure cluster), bottom row, a side view.

The variation of the degree of ordering of the *sp²* carbon structures layout in the domains gives three types of the domains: disordered, weakly-ordered and ordered ones (see Sec. 2). To vary the density of each type of the domain and to cover the expected range of densities of the structured component of the samples we modeled the ensembles of the *sp²* carbon structures with $R_{opt}$ parameter value (see Eq. 1) set for each pair of structures in the ensemble in the following ranges: 0.25 - 0.3 nm, 0.3 - 0.35 nm, 0.35 - 0.4 nm, and 0.4 - 0.45 nm. The total number of variants of atomic density and degree of ordering in a single domain amounts to $N_p = 12$. The total number of the domains $N$ is equal to $N_{str} \cdot N_p = 432$, in the case of mono-structure domains, and $N_p = 12$, in the case of multi-structure domains.

Fig. 4 shows a comparison of experimental and theoretical neutron diffraction patterns for the analyzed samples. A good agreement between the theoretical and experimental curves is achieved for all samples. In the cases of $S_{25}$ and $S_{600}$ samples the theoretical curve obtained in the case of multi-structure domains poorly reproduce the behavior of the experimental curve at $q<10$ nm$^{-1}$, while the curve obtained in the case of mono-structure domains fits the experiment well. For the other samples it is impossible to determine (at least using the diffraction data only) whether the mono-structure domains are present in the sample or not, but taking into account the closeness of the shapes of the considered *sp²* carbon structures (the curved flakes mostly) it is very unlikely. The success in the fitting of the experimental data allows us to estimate the important structural properties of the samples such as the average size and curvature of the *sp²* carbon structures as well as the average density of a structured component and the degree of ordering of structures layout in the sample.

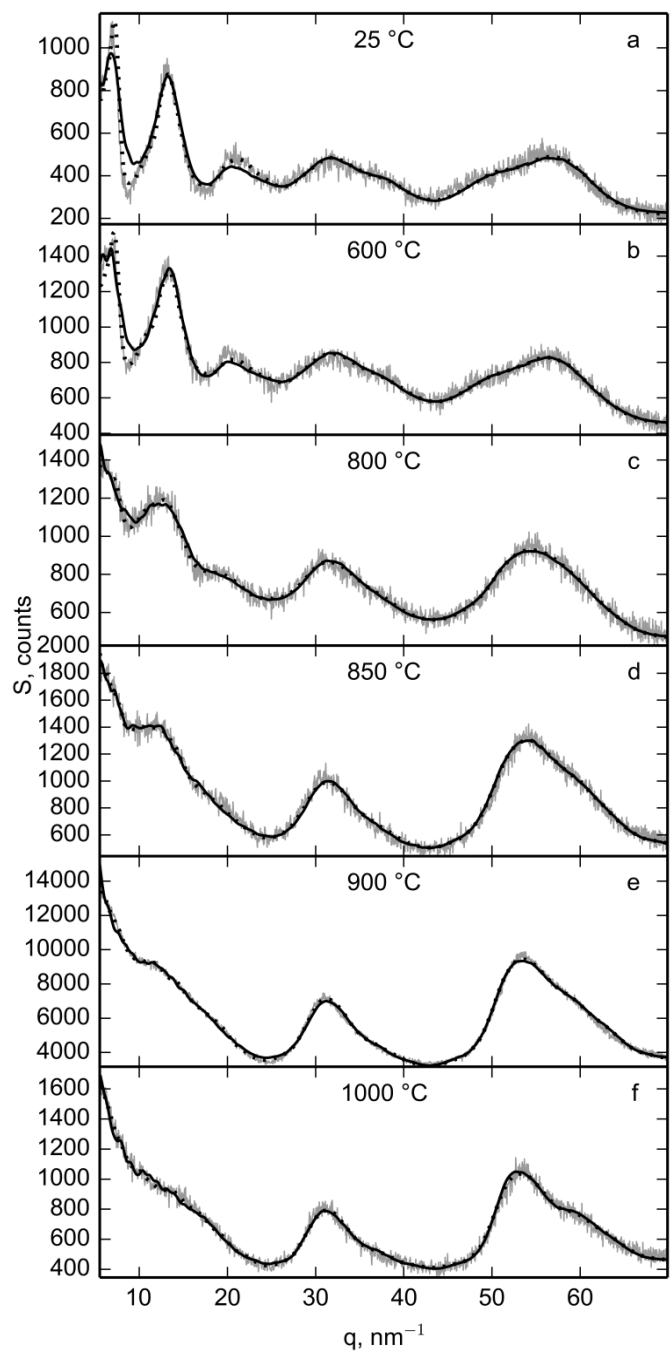

Fig. 4. Comparison of experimental (grey) and theoretical (mono-structure domains, black dotted, and multi-structure domains, black solid) neutron diffraction patterns for various samples annealed at different temperatures ("a" corresponds to the unannealed sample). The exposure time is different for different samples.

The result for the *sp²* carbon structures size estimation is shown in Fig. 5. We divided the structures into five groups: (i) $C_{60}$ fullerene molecules, (ii) small-size flakes with up to 41 atoms, (iii) medium-size flakes, with 41 – 80 atoms, (iv) large-size flakes, with 81 – 120 atoms and (v) extra-large-size flakes, with number of atoms greater than 120. The atomic fraction for the structures from each of the five groups in the structured component of the samples is shown. The fraction of the $C_{60}$ fullerenes decreases monotonically with increasing annealing temperature of the sample, reaching zero for the $S_{1000}$ sample. The fundamental change of the sample's structure occurs at annealing temperature in the range from 600 to 800 °C, while the fraction of the $C_{60}$ fullerenes decreases by a factor of 5. Interestingly, the fraction of the $C_{60}$ fullerenes in the structured component of the $S_{25}$ sample is only 60 %, and small- and medium-size flakes are also present. The large- and extra-large-size flakes are absent in the $S_{25}$ and $S_{600}$ samples. The fraction of the extra-large-size flakes increases monotonically with increasing annealing temperature of the sample, reaching 40% for the $S_{1000}$ sample. The fraction of medium-size flakes is high in all annealed samples varying in the range from 30% to 50%. The fraction of small-size flakes reach maximum value of 43% for the $S_{850}$ sample. The average number of atoms in the $sp^2$ carbon structure is shown in Fig. 6. It is seen that the size of the structures is monotonically increasing with the annealing temperature.

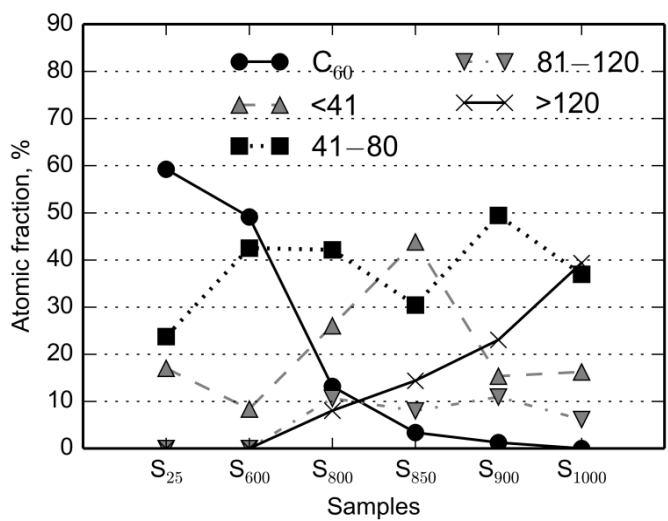

Fig. 5. The atomic fraction for the $sp^2$ carbon structures of various number of atoms in the structured component of six samples of Fig. 4.

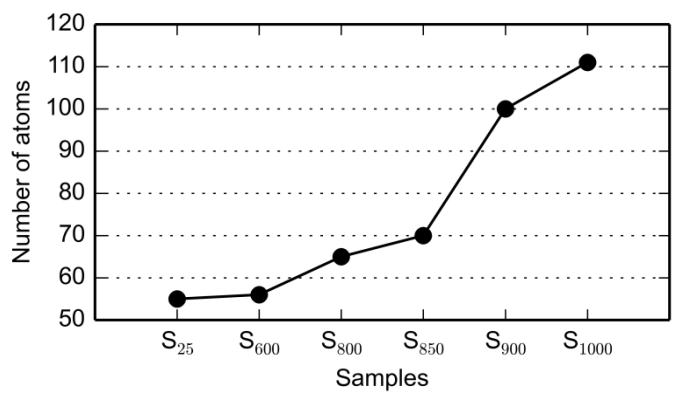

Fig. 6 Average number of atoms of the $sp^2$ carbon structures in the six samples of Fig. 4.

The estimated curvature of the $sp^2$ carbon structures is shown in Fig. 7. We divided the structures into five groups by the degree of their curvature in terms of the curvature (sphere or tube) radius values: (i) 0.35 - 0.4 nm (including $C_{60}$ fullerenes), (ii) 0.4 - 0.6 nm, (iii) 0.6 - 0.8 nm, (iv) 0.8 - 1.3 nm, (v) flat flakes. The atomic fraction of highly curved structures in the structures component of the sample decreases with increasing annealing temperature of the sample, while the fraction of slightly curved structure increases. The fraction of flat flakes is low and exceeds 15 % only for the $S_{25}$ and $S_{1000}$ samples. The number of atoms in the flat flakes in the $S_{1000}$ sample is 7 times higher than that in the $S_{25}$ sample. The average curvature of the $sp^2$ carbon structures defined as the mean of the inverse curvature radius of structures is shown in Fig. 8. The curvature of the structures decreases with increasing annealing temperature of the sample.

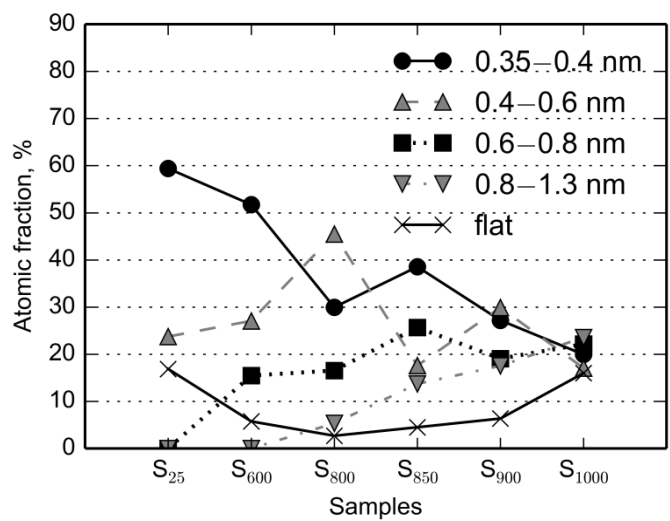

Fig. 7. The atomic fraction for the $sp^2$ carbon structures of various curvature in the six samples of Fig. 4. The curvature is given in terms of the curvature radius that is either a sphere or a tube radius ("flat" corresponds to the flat structures).

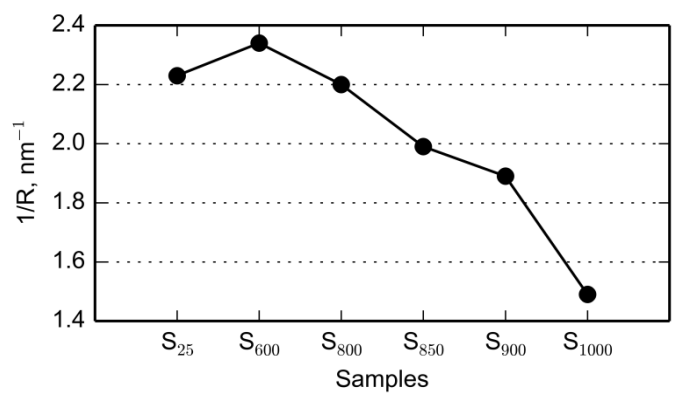

Fig. 8. Average curvature of the $sp^2$ carbon structures in the six samples of Fig. 4 defined as inverse curvature radius (that is either a sphere or a tube radius) of structure.

The accuracy for the recovery of degree of ordering and density of the $sp^2$ carbon structures in the structured component of the sample is lower than that for the recovery of their size and curvature. However it's worth to analyze these parameters as well. We obtain the following results. Fig. 9 shows the fraction for the disordered, weakly ordered and ordered domains of the $sp^2$ carbon structures in the structured component of the samples. The results for the $S_{25}$ and the $S_{600}$ samples are not shown because a high fraction of $C_{60}$ fullerenes (i.e. of highly symmetric structures) does not allow to differentiate the domains by the degree of ordering. The fraction of the disordered domains is decreasing with increasing annealing temperature. For the weakly-ordered and ordered domains the results are not obvious mainly because the domains mostly consist of curved flakes which layout in the domain influences the diffraction patterns lesser than that of the flat flakes. The weakly ordered domains dominate in the $S_{1000}$ sample.

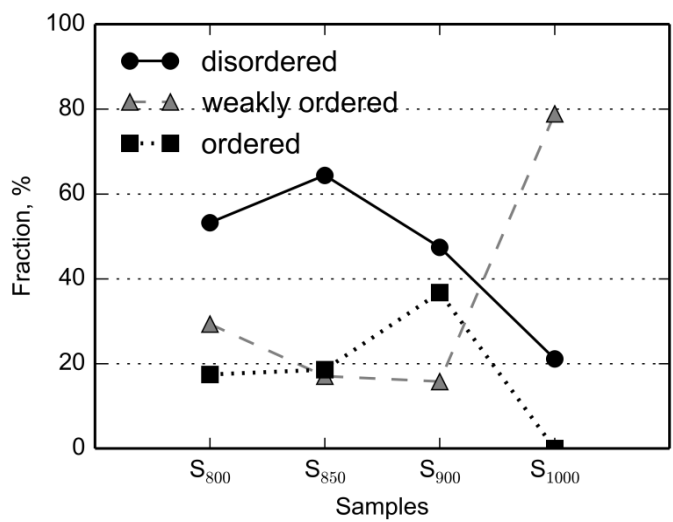

Fig. 9. The fraction for the domains of the $sp^2$ carbon structures of different degree of mutual layout ordering in the six samples of Fig. 4.

We divided the domains into three groups by their density: (i) 1.8 – 2.3 g/cm$^3$, (ii) 1.4 – 1.8 g/cm$^3$ and (iii) 1.0 – 1.4 g/cm$^3$. Fig. 10 shows the fraction for the domains within these groups for the density of a structured component of the samples. The results for the $S_{25}$ and $S_{600}$ samples are shown in the case of mono-structure domains. Significant fraction (18 %) of the domains with density of 1.8 – 2.3 g/cm$^3$ is present only in the $S_{25}$ sample. The fraction of the domains with density of 1.0 – 1.4 g/cm$^3$ is increasing with increasing annealing temperature, reaching the value of 100 % for the $S_{1000}$ sample, while the fraction of domains with density of 1.4 – 1.8 g/cm$^3$ is decreasing to zero.

The average density of the carbon structured component in the samples is shown in Fig. 11. It decreases with increasing annealing temperature from 1.48 g/cm$^3$ for the $S_{25}$ sample to 1.21 g/cm$^3$ for the $S_{1000}$ sample.

Another parameter that we can restore is the intermolecular distance between the $C_{60}$ fullerenes in the $S_{25}$ sample. It equals to 0.33 – 0.34 nm, that is lower than intermolecular distance in the crystalline $C_{60}$ (0.3 nm).

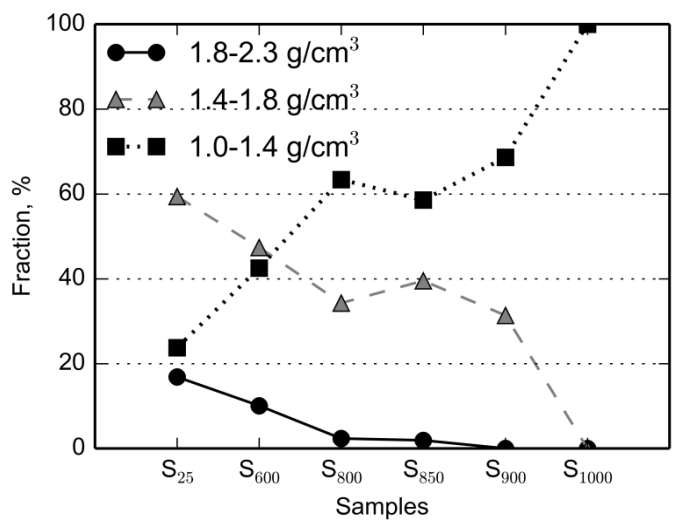

Fig. 10. The fraction for the domains of different density in the structured component of six samples of Fig. 4.

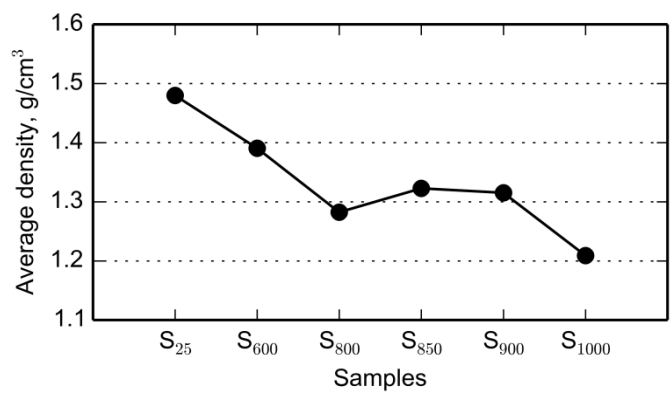

Fig. 11. The average density of the carbon structured component in the six samples of Fig. 4.

## 5. Conclusions

We suggested a method for estimation of structural properties of amorphous fullerene and its derivatives produced by vacuum annealing. The main features of the method are as follows:
  a) fitting of the neutron or x-ray powder diffraction patterns for scattering wave vector's modulus $q$ in the range from few units to several tens of inverse nanometers;
  b) a structured component of a sample is described with a limited number, $N_{str}$, of structural blocks of a limited number of atoms, $N_{atom}$;
  c) the above structures in a sample are packed heterogeneously, in the domains of a linear size > 10 nm, with various average atomic density and various degree of ordering of structures mutual orientation (for $q$ in the above-mentioned range, the interference of diffraction by separate domains may be neglected);
  d) packing of structures is modeled using the Rigid Body Molecular Dynamics with variable parameter of pair interaction of atoms in the neighboring rigid-body structures.

The method is applied to interpreting the data of neutron diffraction by an amorphous fullerene annealed in a vacuum at 600, 800, 850, 900 and 1000 °C. The set of structural blocks includes the $sp^2$ carbon structures of various curvature and size, for $N_{str} = 36$ and $N_{atom} = 14 \div 285$. The variation of domains by their average atomic density and degree of mutual orientation ordering of structures has 12 variants. We quantify structural properties of the samples in terms of the average size and curvature of the $sp^2$ carbon structures, and analyze sensitivity of results to the layout of these structures in the domains. The implementation of item (a) is done using the optimization procedure and the respective RESTful web-services [6] created in the Mathcloud distributed environment [7]. The numerical simulations of item (d) are carried out using the high-performance computational cluster of the NRC "Kurchatov Institute".

We found that the atomic fraction of the $C_{60}$ fullerenes decreases monotonically with increasing annealing temperature of the sample, reaching zero for the sample annealed at 1000 °C. A substantial change of the sample's structure occurs at annealing temperature in the range from 600 to 800 °C - in particular, the fraction of the $C_{60}$ fullerenes decreases by a factor of 5. The structured component of the samples annealed at 800 and higher temperatures is dominated by the curved $sp^2$ carbon flakes of 60 – 110 atoms. The average number of atoms in the flakes is monotonically increasing with the annealing temperature, while the curvature is decreasing. The density of the structural component of the sample is decreasing with increasing annealing temperature, while mutual orientation of the flakes in the domains becomes more ordered.

**Acknowledgments**

The authors are grateful to A.P. Afanasiev, for the support of collaboration between the NRC "Kurchatov Institute" and the Center for GRID-Technologies and Distributed Computing (http://dcs.isa.ru) of the Institute for Information Transmission Problems (Kharkevich Institute) of Russian Academy of Science, V.A. Somenkov, for a guidance in the experimental part of the

work, O.V. Sukhoroslov, for technical support of MathCloud computing infrastructure, Y.V. Zubavichus, for helpful discussion of results.

This work is supported by the Russian Foundation for Basic Research (projects RFBR #12-07-00529-a, #13-07-00987-a, and #13-02-00208-a). The experimental work was performed using the equipment of Unique Scientific Facility "Research reactor IR-8" supported by the Russian Ministry of Science and Education (project code RFMEFI61914X0002). The most of calculations were carried out using the computational resources of the Multipurpose Computational Complex of the NRC "Kurchatov Institute" (http://computing.kiae.ru/).